\documentclass[%
reprint,
amsmath,amssymb,
prb,
]{revtex4-2}

\usepackage{graphicx}
\usepackage{dcolumn}
\usepackage{bm}
\usepackage{subfigure}
\usepackage{booktabs}
\usepackage[colorlinks=true, letterpaper=ture, pdfstartview=FitV, linkcolor=blue, citecolor=blue, urlcolor=blue]{hyperref}

\begin{document}

\preprint{APS/123-QED}

\title{Large-scale Atomistic Simulation of Quantum Effects in SrTiO$_3$ from First Principles}

\author{Hongyu Wu$^{1,2}$}
\thanks{These authors contributed equally to this work.}
\author{Ri He$^{1}$}
\thanks{These authors contributed equally to this work.}
\author{Yi Lu$^{3,4}$}
\author{Zhicheng Zhong$^{1,5}$}
\thanks{zhong@nimte.ac.cn}

\affiliation{\\$^1$CAS Key Laboratory of Magnetic Materials and Devices $\&$ Zhejiang Province Key Laboratory of Magnetic Materials and Application
Technology, Ningbo Institute of Materials Technology and Engineering, Chinese Academy of Sciences, Ningbo 315201, China}

\affiliation{$^2$College of Materials Science and Opto-Electronic Technology, University of Chinese Academy of Sciences, Beijing 100049, China}

\affiliation{$^3$National Laboratory of Solid State Microstructures and Department of Physics, Nanjing University, Nanjing 210093, China}

\affiliation{$^4$Collaborative Innovation Center of Advanced Microstructures, Nanjing University, Nanjing 210093, China}

\affiliation{$^5$China Center of Materials Science and Optoelectronics Engineering, \\
University of Chinese Academy of Sciences, Beijing 100049, China}


\date{\today}

\begin{abstract}
Quantum effects of lattice vibration play a major role in many physical properties of condensed matter systems, including thermal properties such as specific heat, structural phase transition, as well as phenomena such as quantum crystal and quantum paraelectricity that are closely related to zero-point fluctuations. However, realizing atomistic simulations for realistic materials with a fully quantum-mechanical description remains a great challenge. Here, we propose a first-principle strategy for large scale Molecular Dynamics simulation, where high accuracy force field obtained by Deep-Potential (DP) is combined with Quantum Thermal Bath (QTB) method to account for quantum effects. We demonstrate the power of this DP+QTB method using the archetypal example SrTiO$_3$, which exhibits several phenomena induced by quantum fluctuations, such as the suppressed structure phase transition temperature, the quantum paraelectric ground state at low temperature and the quantum critical behavior $1/T^2$ law of dielectric constant.
Our DP+QTB strategy is efficient in simulating large scale system, and is first principle. More importantly, quantum effects of other systems could also be investigated as long as corresponding DP model is trained. This strategy would greatly enrich our vision and means to study quantum behavior of condensed matter physics.
\end{abstract}

\maketitle


\section{Introduction}
Atomic nuclei can be simply treated as classical particles with certain position and velocity,
but considering the quantum mechanical nature of the atomic nuclei, their quantum effects need to be considered in some cases\cite{kittel1996introduction}. Based on the hypothesis of energy quantization and Bose-Einstein statistics, collective excitation of atoms can be viewed as phonon. Based on theory of phonon, we can calculate thermal conductivity, electrical conductivity and specific heat of solid, these are connected to excitation. Particularly at ground state, zero-point energy and uncertainty principle induce zero-point fluctuations of atomic position. The zero-point fluctuation is crucial for an accurate description of phenomena such as quantum crystal, high pressure superconductivity, quantum nature of hydrogen/water, and quantum paraelectricity\cite{RevModPhys.89.035003,errea2020quantum,Morrone2008,Li2011a,PhysRevB.19.3593,rowley2014ferroelectric}.
Although phonon language have success in describing low temperature, harmonic and simple systems in \emph{q}-space, for many problems that urgently need to know the atomistic position information in real space and time, such as point deffects, dislocations, grain boundary and free surface, it is highly desirable to introduce quantum effects in the atomistic simulation of large-scale systems.

The widely accepted method to incorporate quantum mechanics in atomistic simulation is path integral molecular dynamics (PIMD)\cite{Marx1994,marx1996ab}. In PIMD, the nuclei part is mapped to a system composed of several virtual particles connected by a harmonic potential, and its effective Hamiltonian is derived from Feynman path integral. The PIMD methods have two limitations in practice. The first is the lack of accurate force field. Without accurate force field, the major part of calculation is not reliable, even adding quantum correction will not give realistic results\cite{Miller2005}. Secondly, extremely expensive calculation costs hinder the practical use of PIMD. PIMD is very time consuming, data applicability ( e.g. space scale of simulation cell, time scale and time step of simulation ) is usually reduced in related work\cite{Ceperley1995}. When it comes to statistics related issues, such as thermodynamic properties, phase diagrams, etc., the results of PIMD are difficult to achieve statistical equilibrium due to scale constraints. Therefore, it is necessary to develop a feasible and ab-initio numerical scheme to describe the quantum effects in large scale atomistic simulation.

We propose a Deep-Potential (DP) + Quantum Thermal Bath (QTB) strategy which can provide quantum correction and yet is capable of large scale atomistic dynamic simulation with density-functional-theory (DFT) precision.
DP is a machine learning method which can produce accurate force fields of molecular dynamics (MD) by sampling the results of DFT\cite{zhang2018deep,Zhang2018}. Some early works in this field have showcased that DP is capable of describing large systems with high accuracy\cite{PhysRevLett.126.236001,PhysRevB.105.064104}.
QTB is a method that successfully preserving the features of quantum statistics in MD\cite{dammak2009quantum}. Its core idea is rewriting the fundamental Newtonian equations in MD into Langevin-like equations with colored noise that incorporate quantum statistics. The computation cost of DP+QTB is slightly higher than classical MD, so in principle can provide results with large space and time scale, from which we can obtain sufficient and intuitive physical information.

\begin{figure}[t]
\centering
\includegraphics[width=\columnwidth]{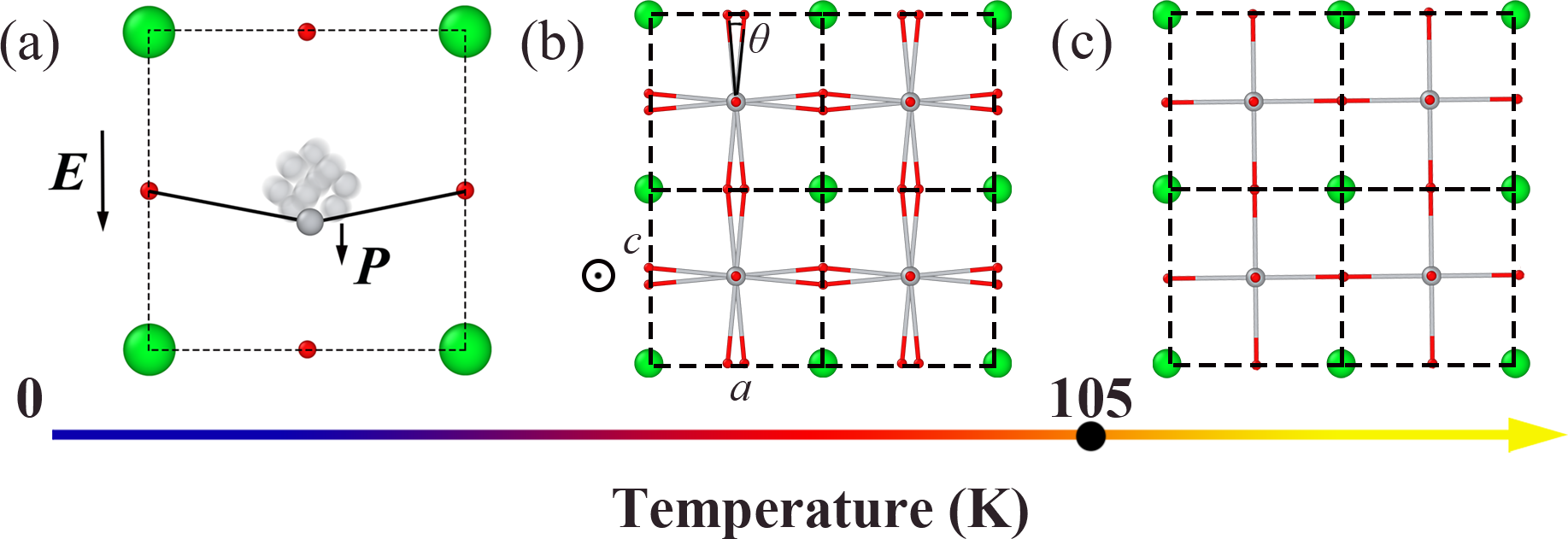}
\label{SrTiO$_3$schematic}
\caption{Schematic diagram of temperature dependent structure of SrTiO$_3$ (a) the response of Ti atom to electric field at zero temperature, the linear response corresponds to the classical case, and multiple transparent atomic trajectories correspond to the quantum zero point fluctuation. (b) tetragonal phase, where \emph{a} and \emph{c} are the short and long axis, $\theta$ is the rotation angle of TiO$_6$ octahedrain each unit cell. (c) cubic phase, where all lattice constant are equal and $\theta$ is zero. }
\label{STOScheme}
\end{figure}

To demonstrate the power of DP+QTB, we focus on the strontium titanate (SrTiO$_3$). SrTiO$_3$ as a representative ABO$_3$ perovskite has been reported to exhibit rich physical properties such as high-mobility electron gas\cite{Ohtomo2004}, flexoelectricity\cite{Zubko2007}, optically strained metastable ferroelectricity \cite{Nova2019}, superconducting dome\cite{Edge2015}, and multiple structural instabilities\cite{PhysRevLett.74.2587}. What makes it a suitable candidate for testing the DP+QTB method is that its atomic structure has several interesting properties. At low temperature, the quantum atomic fluctuation stabilizes a paraelectric phase with a high dielectric constant instead of ferroelectric phase in the classical case (Figure~\ref{STOScheme} (a)), a phenomenon known as the quantum paraelectricity\cite{PhysRevB.19.3593,rowley2014ferroelectric,chandra2017prospects,Shin2021}. With increasing temperature, SrTiO$_3$ also shows a tetragonal to cubic structural phase transition at about 105 K (Figure~\ref{STOScheme} (b) $\&$ (c)).

In this work, we carried out massive DP+QTB simulations of SrTiO$_3$ up to 135000 atoms per simulation cell. Compared with classical simulation results, the DP+QTB show several qualitative improvement. First, zero-point expansion of the SrTiO$_3$ is observed, which compensates the underestimation of lattice constants in most classical calculations\cite{Wexler2019}. Second, the tetragonal-to-cubic phase transition temperature of SrTiO$_3$, which is vastly overestimated in classical simulations, is now much closer to the real transition temperature. Finally, in the quantum critical region at low temperature, the 1/T$^2$ law of the dielectric constant $\varepsilon(T)$ is observed, in agreement experimental results, analytical phenomenological model\cite{PhysRevB.19.3593,rowley2014ferroelectric}.

\section{computational methods}

\subsection{Deep Potential of SrTiO$_3$}
DP is a machine-learning-based method that aims to provide highly applicable and high-accuracy interatomic interaction potentials\cite{zhang2018deep}. The core idea of DP model is to fit the energy and force of different configurations of certain material through a deep neural network, and its learning samples are a large number of structural configurations ( usually these configurations have no more than 50 atoms ) and corresponding high-precision DFT reults (energy and force). For a well-trained model, given any corresponding configuration (include those $\bf{NOT}$ in the training data and supercells with a large number of atoms), and its total energy and the force of each atom can be solved with minimal computational cost and at DFT-level accuracy. Thus, DP model can be applied as a force field in MD simulations.

To obtain the DP model of SrTiO$_3$ we used in this work, $\sim$2000 candidate configurations which contain information of SrTiO$_3$ were selected for training dataset. The accuracy of DP model of SrTiO$_3$ is confirmed in our previous work\cite{PhysRevB.105.064104}. The DP predicted energy difference of tetragonal and cubic phase is 1.008 meV/atom, which is close to the DFT result $\sim$1.1 meV/atom. Such subtle variation is crucial for the tetragonal-to-cubic phase transition in SrTiO$_3$. Other properties such as lattice constant, polarization, elastic constants and phonon dispersion are tested and all agree with DFT calculations. For more information about DP and generation of dataset, please refer to the Supplementary Materials (Discussion S1 and Figure S1)\cite{Supplementary}.

\subsection{Quantum Thermal Bath}
MD simulations are based on Newtonian mechanics equations and classical statistical theory, therefore only produce results that conform to classical behavior\cite{Plimpton1995}. QTB is based on the quantum mechanical fluctuation-dissipation theorem. Instead of treating the atomic nuclei as independent point-like classical particles, it introduces associated random force and friction term into the equation, constituting a quantum thermal bath\cite{dammak2009quantum}. The equation of motion is modified to Langevin-like equation
\begin{equation}
m_i\ddot{r}_{i\alpha}=f_{i\alpha}+R_{i\alpha}-m_i\gamma\dot{r}_{i\alpha},
\end{equation}
where $m_i$ is the mass of ith atom, $r_{i\alpha}$ and $f_{i\alpha}$ are the $\alpha$ (1, 2 or 3) components of the position and the interatomic force exerted by all the other atoms. $R_{i\alpha}$ is Gaussian random force and $\gamma$ is effective frictional coefficient. The random force spectrum is a colored noise, its power spectral density is related to $\gamma$ by the quantum mechanical fluctuation-dissipation theorem\cite{PhysRev.83.34}
\begin{equation}
I_{R_{i\alpha}R_{j\beta}}(\omega)=2m_i\gamma\delta_{ij}\delta_{\alpha\beta}\theta(|\omega|,T),
\end{equation}
where
\begin{equation}
\theta(\omega,T)=\hbar\omega[\frac{1}{2}+\frac{1}{exp(\frac{\hbar\omega}{k_BT})-1}],
\label{PSD}
\end{equation}
$\delta_{ij}$ and $\delta_{\alpha\beta}$ are the Kronecker symbol and $k_B$ is the Boltzmann constant. Please refer to the Supplementary Materials (Discussion S2 and Figure S2) for details of atomic forces in MD without and with QTB\cite{Supplementary}. It is crucial to emphasize that in eq.\eqref{PSD}, the first term $\frac{1}{2}\hbar\omega$, namely the zero-point energy, has no analog in classical theories. Further, the numerical techniques related to QTB were improved based on simple harmonic oscillators\cite{barrat2011portable}, and QTB can be easily manipulated in practice and is independent of the studied system. Previous works have demonstrated that QTB+MD can produce results in good agreement with experiments or PIMD results\cite{dammak2009quantum,barrat2011portable,PhysRevLett.107.198902} with computation complexity comparable to classical MD.

\subsection{Dielectric Constant Calculation}
By adding effective electric field along axis of SrTiO$_3$ in MD simulations, we can obtain the dielectric constant via linear response  $\bf{P}=\varepsilon\varepsilon_0\bf{E}$. The local pseudocubic cell polarization $\mathbf{P}$ can be calculated by atomic displacements with respect to the referenced cubic phase multiplied by the Born effective charges:
\begin{equation}
\mathbf{P}=\sum\mathbf{Z}_i^*\mathbf{u}_i.
\end{equation}
The components of Born effective charge tensors along the out-of-plane direction were obtained by DFT calculations: $Z_{Sr}^*$= 2.54, $Z_{Ti}^*$= 7.12, $Z_{O1}^*$= -5.66, $Z_{O2}^*$= -2.00, where $O1$ denotes the oxygen atom in the SrO layer, and $O2$ denotes the oxygen atom in the TiO$_2$ layer.

\section{results and discussions}
\subsection{Thermodynamic Properties of SrTiO$_3$}
\begin{figure*}[htbp]
	\centering
	\subfigure{
		\includegraphics[width=\columnwidth]{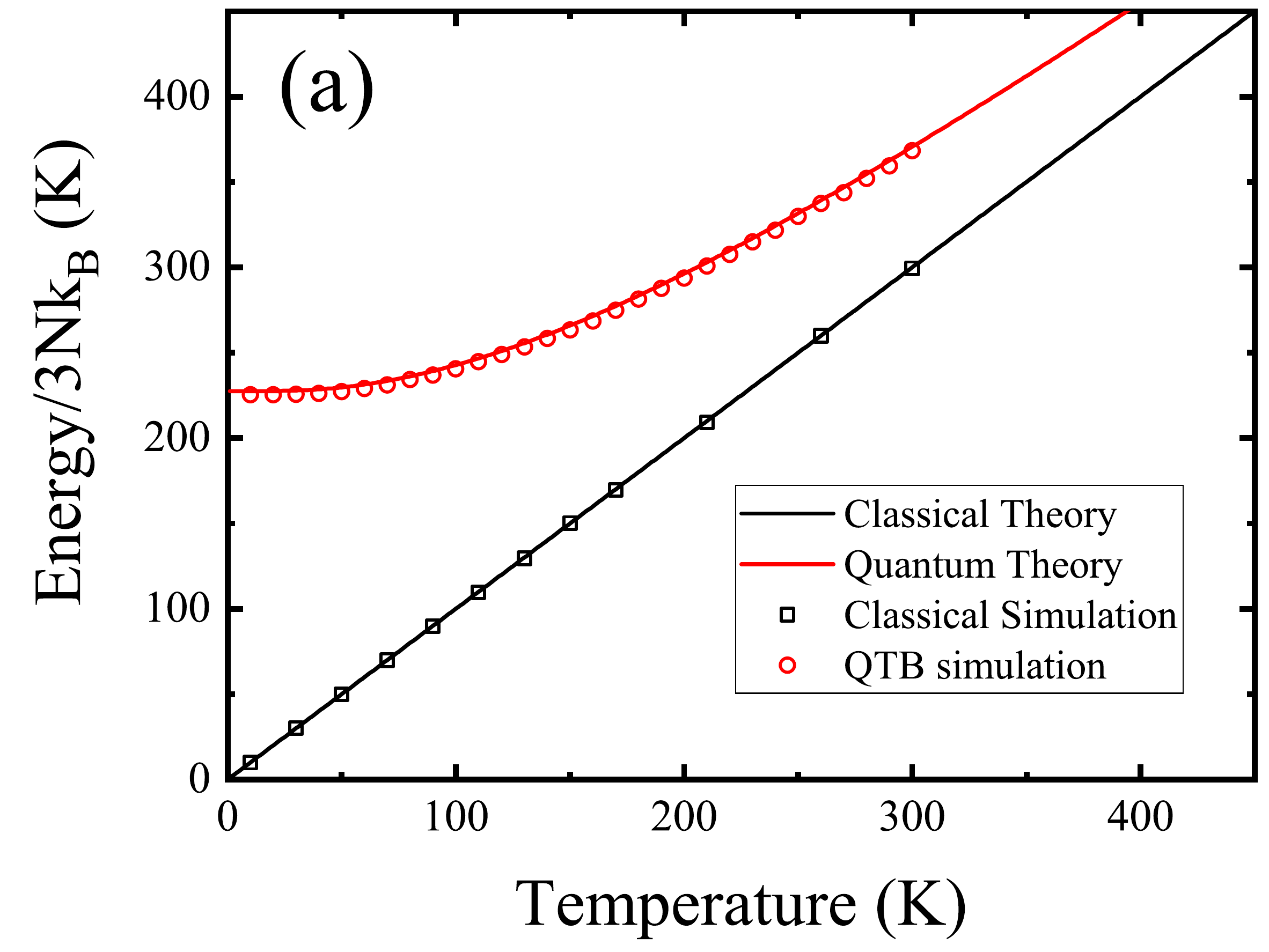}
		\label{TTvsFT}
	}
	\subfigure{
		\includegraphics[width=\columnwidth]{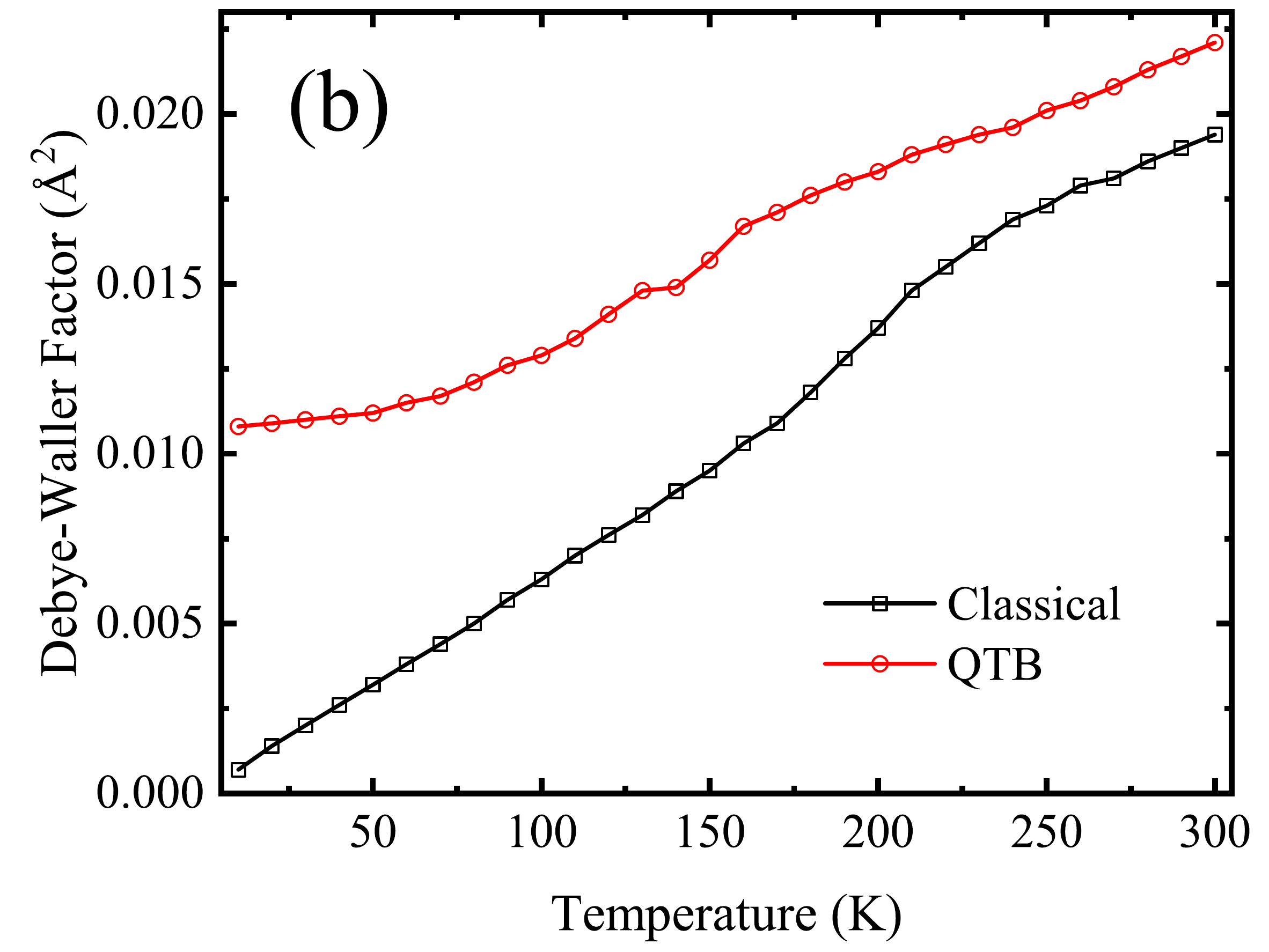}
		\label{DWF}
	}
	\caption{Comparison of thermodynamic properties between classical and quantum simulations. (a) Average energy per atom in SrTiO$_3$ as a function of temperature. (b) Debye-Waller factor, or average mean square displacement, in SrTiO$_3$ as a function of temperature.}
	\label{Thermal}
\end{figure*}

The thermodynamic properties are affected by the quantum effects of lattice. By comparing the classical and quantum theoretical and simulation results, we can on the one hand test the validity of the DP+QTB method, and on the other hand characterize the strength of the quantum effect at any temperature. We first compute the evolution of energy as a function of temperature and show it in the form of $\frac{E}{3Nk_B}$ for a more intuitive display. As shown in Figure~\ref{TTvsFT}, the black line and dots represent classical harmonic oscillator theory and classical MD simulation results, respectively. The energy of SrTiO$_3$ in classical case are equal to $3Nk_BT$ at any temperature, as dictated by the equipartition theorem of classical Boltzmann statistics.

The DP+QTB simulation results are shown in red dots in Figure~\ref{TTvsFT}. At 0 K, the energy of system is not 0, which is the zero-point energy. At finite temperature, the increase of energy with temperature rise in the quantum case is not as high as that in the classical case, which is due to the energy level gap of quantum excitation, and only part of the phonons contribute to the excitation. At room temperature (300 K), the difference between the classical and quantum results is still about 70K in Figure~\ref{TTvsFT}, which indicates that even at room temperature, which was previously recognized as a classical scenario, SrTiO$_3$ is still far from the classical limit.
The lattice constants and elastic constants of SrTiO$_3$ under classical and quantum conditions to measure the strength of quantum effects, as shawn in Table~\ref{lattice}. Compared to classical MD 0K or according to the definition of DP model, it essentially represents the zero temperature result predicted by DFT, quantum effects are mapped to expansion of lattice constants and reduction of rotation angle $\theta$. As for the elastic constants, quantum effects only slightly affects them.

Based on the theory of quantum harmonic oscillator, the temperature dependent energy per atom is given by:
\begin{equation}
E(T)=\int_{0}^{\infty}g(\omega)\theta(\omega,T)d\omega,
\end{equation}
where $g(\omega)$ is the normalized phonon density of states of SrTiO$_3$, as drawn in the red line in Figure~\ref{TTvsFT}. The QTB results are able to recover the quantum harmonic approximation and show convergence to the classical limit at high temperatures.

\begin{table}[b]
\centering
\caption{Lattice constants and elastic constants of SrTiO$_3$ simluated by classical MD and DP+QTB}
\label{lattice}
\begin{ruledtabular}
\begin{tabular}{lccccc}
\textrm{}&
\multicolumn{2}{c}{\textrm{0K}}&
\textrm{}&
\multicolumn{2}{c}{\textrm{300K}}\\ \cmidrule(lr){2-3} \cmidrule(l){5-6}
\textrm{} & \textrm{Classical} & \textrm{QTB} & \textrm{} & \textrm{Classical} & \textrm{QTB} \\
\colrule
a/$\textrm{\AA}$      & 3.888        & 3.904   &  & 3.913         & 3.915    \\
c/$\textrm{\AA}$      & 3.910        & 3.914   &  & 3.913         & 3.915    \\
$\theta$/°  & 5.49         & 4.75    &  & 0             & 0        \\
C$_{11}$/GPa & 363.18       & 349.78  &  & 340.45        & 335.45   \\
C$_{12}$/GPa & 106.85       & 100.33  &  & 84.84         & 80.41    \\
C$_{44}$/GPa & 100.89       & 101.55  &  & 90.34         & 88.79    \\
\end{tabular}%
\end{ruledtabular}
\end{table}

To measure the strength of the quantum effect, we define zero-point temperature
$T_{zero}=\frac{1}{k_B}\int_{0}^{\infty}g(\omega)\frac{1}{2}\hbar\omega d\omega$, it represents the average energy per atom of zero-point fluctuation in form of temperature.
$T_{zero}$ is about 230K which is contributed by the zero-point energy of SrTiO$_3$, such a large difference from the classical results suggests that contribution of quantum effects is substantial. We can also determine $T_{zero}$ in other materials to calibrate the strength of quantum effects in the material, $T_{zero}$ of some typical materials are shown in Table~\ref{Tzero}. According to the simple harmonic approximation $\omega=\sqrt{\frac{K}{M}}$, where $\omega$ and $K$ are effective frequency and spring constant, $M$ is the mass of atom, we can qualitatively analyze $T_{zero}$. For metals with weak bonding interactions, their effective spring constant $K$ is weaker, and their $\omega$ is smaller, so their $T_{zero}$ is generally smaller. On the contrary, in covalent materials, their $K$ is larger, so their $T_{zero}$ is generally larger, which indicates that quantum effects are also strong in these materials. In some special materials, such as LaH$_{10}$ high pressure hydrogen-rich superconducting material, its effective spring coefficient will be very high because of the special environment under high pressure, and hydrogen is light element, so the $T_{zero}$ of LaH$_{10}$ is abnormally high. It can be seen from $T_{zero}$ that it is not limited to low temperature and light elements, the quantum effects of many systems are very significant, which also shows the necessity of the quantum effect of MD simulation.
\begin{table}[b]
\centering
\caption{$T_{zero}$ of some typical materials (K)}
\label{Tzero}
\begin{ruledtabular}
\begin{tabular}{lcclcclc}
\multicolumn{2}{c}{\textrm{Metal}} & \multicolumn{1}{l}{\textrm{}} & \multicolumn{2}{c}{\textrm{Covalent}}  & \multicolumn{1}{l}{\textrm{}} & \multicolumn{2}{c}{\textrm{Others}} \\
\colrule
Al         & 148.0        &                      & B           & 507.9                     &                      & Al$_2$O$_3$      & 351.4   \\
Be         & 346.0        &                      & C(Diamond)  & 697.8                     &                      & LaH$_{10}$       & 818.5   \\
Ca         & 81.6         &                      & C(Graphene) & \multicolumn{1}{l}{666.8} & \multicolumn{1}{l}{} & MgO              & 272.3   \\
Cu         & 122.4        &                      & GaAs        & 122.6                     &                      & MoS$_2$          & 208.0   \\
Fe         & 168.2        &                      & Si(Diamond) & 234.6                     &                      & Na$_2$SiO$_3$    & 288.1   \\
Ta         & 86.7         &                      & SiO$_2$     & 398.5                     &                      & NaCl             & 102.9   \\
Zr         & 90.5         &                      & SrTiO$_3$   & 227.5                     &                      & ZrW$_2$O$_8$     & 270.3   \\
\end{tabular}
\end{ruledtabular}
\end{table}

Similar results can also be obtained from the Debye-Waller factor, or the average mean square displacements of the atom. As shown in Figure~\ref{DWF}, the position fluctuations of the atom in classical case are only contributed by thermal fluctuation, which changes linearly with temperature and atoms freeze at 0 K. In the quantum case, quantum fluctuation of the atomic position appears at 0 K, suggesting that quantum effects bring about the zero-point motion of 0.0076 $\textrm{\AA}^2$, which is contrary to our common sense. The amplitude of zero-point motion are comparable to thermal flucuations at about 150K. And it also returns to the classical limit at high temperature.

\begin{figure}[t]
\centering
\includegraphics[width=\columnwidth]{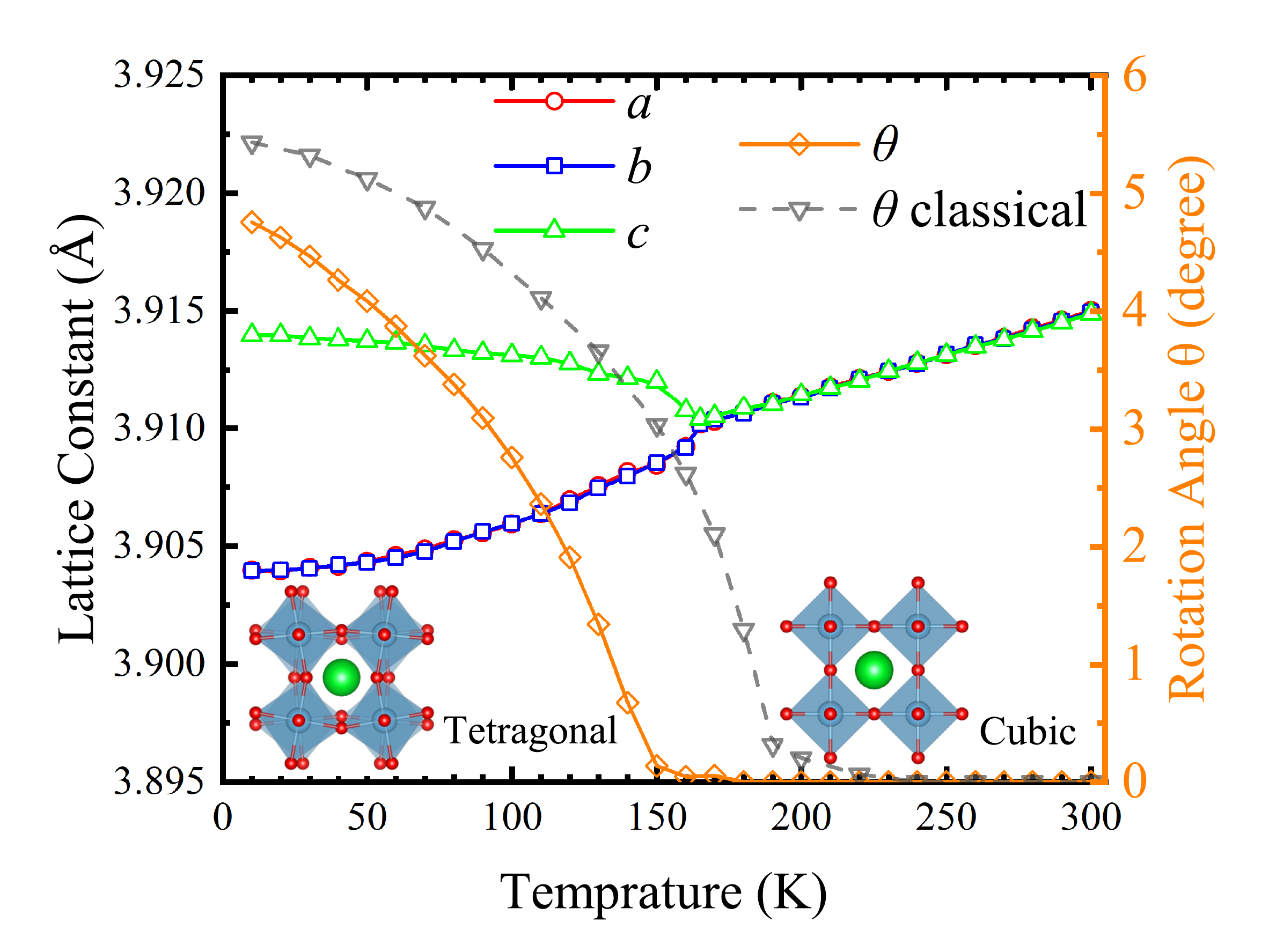}
\caption{Average lattice constants and rotation angle of the TiO$_6$ octahedra around [001] ($\theta$) for SrTiO$_3$ as a function of temperature determined from DP+QTB simulations.}
\label{Trans}
\end{figure}

Since the quantum effects have such great influence on the lattice constants and atomic displacement, it naturally affects structural phase transition. SrTiO$_3$ has a structural phase transition from tetragonal to cubic (Figure~\ref{STOScheme} (b) $\&$ (c)), when the temperature goes up, both the tetragonal distortion (\emph{c/a}) and the rotation angle $\theta$ of TiO$_6$ octahedra around [001] decreases with the temperature increases. In our previous work\cite{PhysRevB.105.064104}, we have already apply DP model of SrTiO$_3$ and classical MD to simulate the phase transition of SrTiO$_3$. The classical MD results show that rotation angle $\theta$ near 0K is 5.48° which is larger than experimental result\cite{Mueller1971}. And the phase transition temperature is about 200K, which is also much higher than experimental result 105K\cite{Mueller1971}. The temperature-dependent lattice constants and rotation angle simulated by DP+QTB are shown in Figure~\ref{Trans}. The $\theta$ decrease to 0 at about 160K, the lattice constants become consistent also around 160K. Anyway, it is clear that DP+QTB successfully introduces quantum effects of lattice vibration and to some extent more reasonably predicts the phase transition temperature of SrTiO$_3$.

Although $T_{zero}$ of SrTiO$_3$ is 227.4 K which is higher than transition temperature simulated by classical MD 200 K, results near 0K ($c/a \neq 1$ and $\theta\neq0$) indicate that the tetragonal phase is still stable. This means that if the zero-point energy is viewed as classical temperature, then it is not simply an increase in temperature, in other words, it is not simply to shift the corresponding temperature. The lattice constants of two short axis (a and b) have a similar behavior to Debye-Waller factor near 0K. So the reason for the decrease of the phase transition temperature compared to classical result is that the quantum fluctuations which exist at finite temperature is remapped, leading to the increase of average displacements and average kinetic energy of atoms, making it easier for atoms to cross the potential barrier to undergo structural transition around 160K. Although DP+QTB still overestimates transition temperature compared to the experimental results. Possible reason is from accuracy of DFT, the energy difference between two phases is about 1meV/atom, we test this energy difference using different exchange correlation function, the virance of these results suggest that it is difficult for DFT to find the exact energy difference. And since this energy difference is critical to determine the transition temperature, the DP model based on the DFT results will also misestimates the transition temperature.

\subsection{Dielectric Constant of SrTiO$_3$}
\begin{figure}[t]
\centering
\includegraphics[width=\columnwidth]{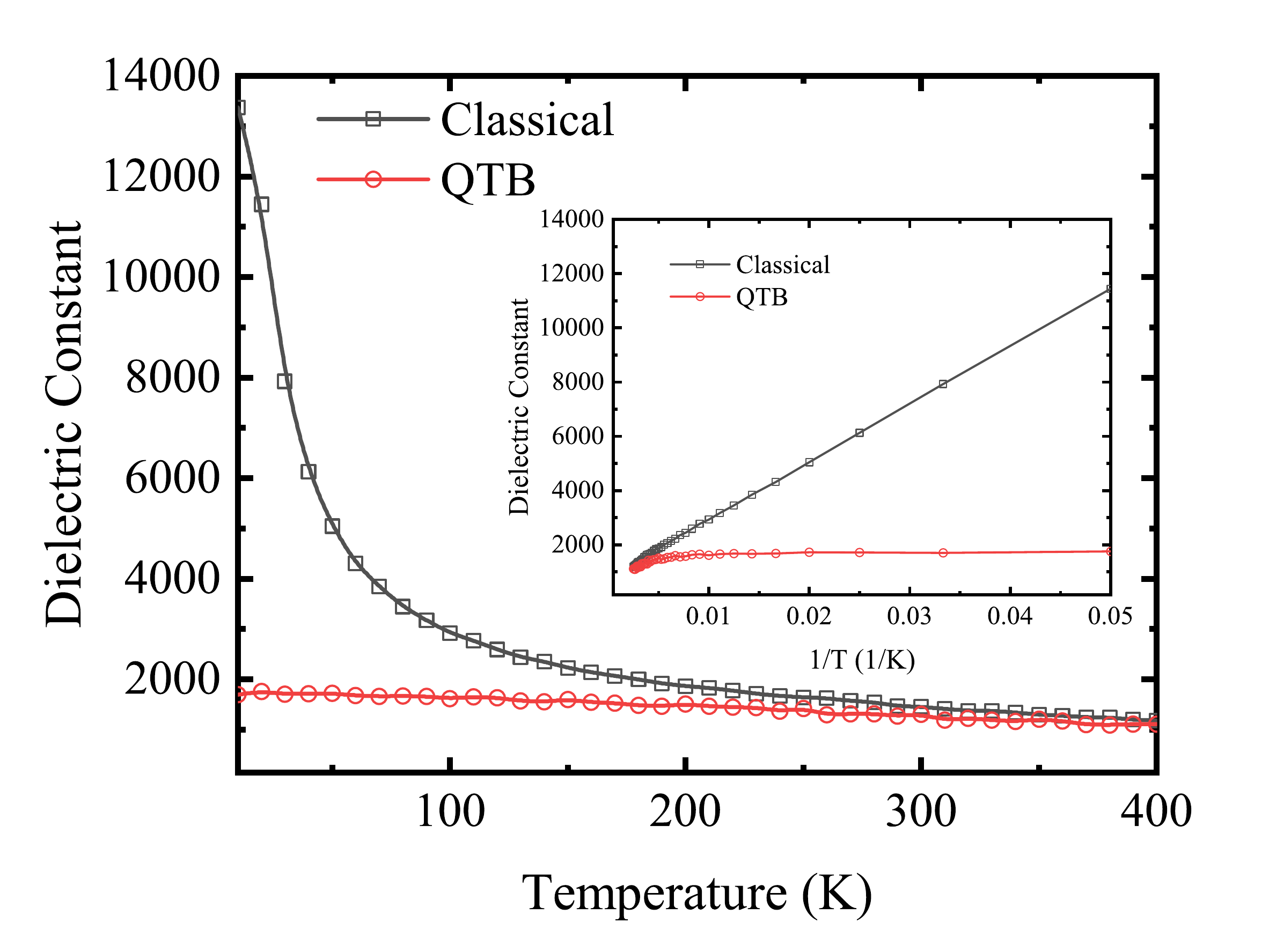}
\caption{Dielectric constant simulated by Classical and QTB simulations.}
\label{ZeroP}
\end{figure}

\begin{figure}[t]
	\centering
	\subfigure{
		\includegraphics[width=\columnwidth]{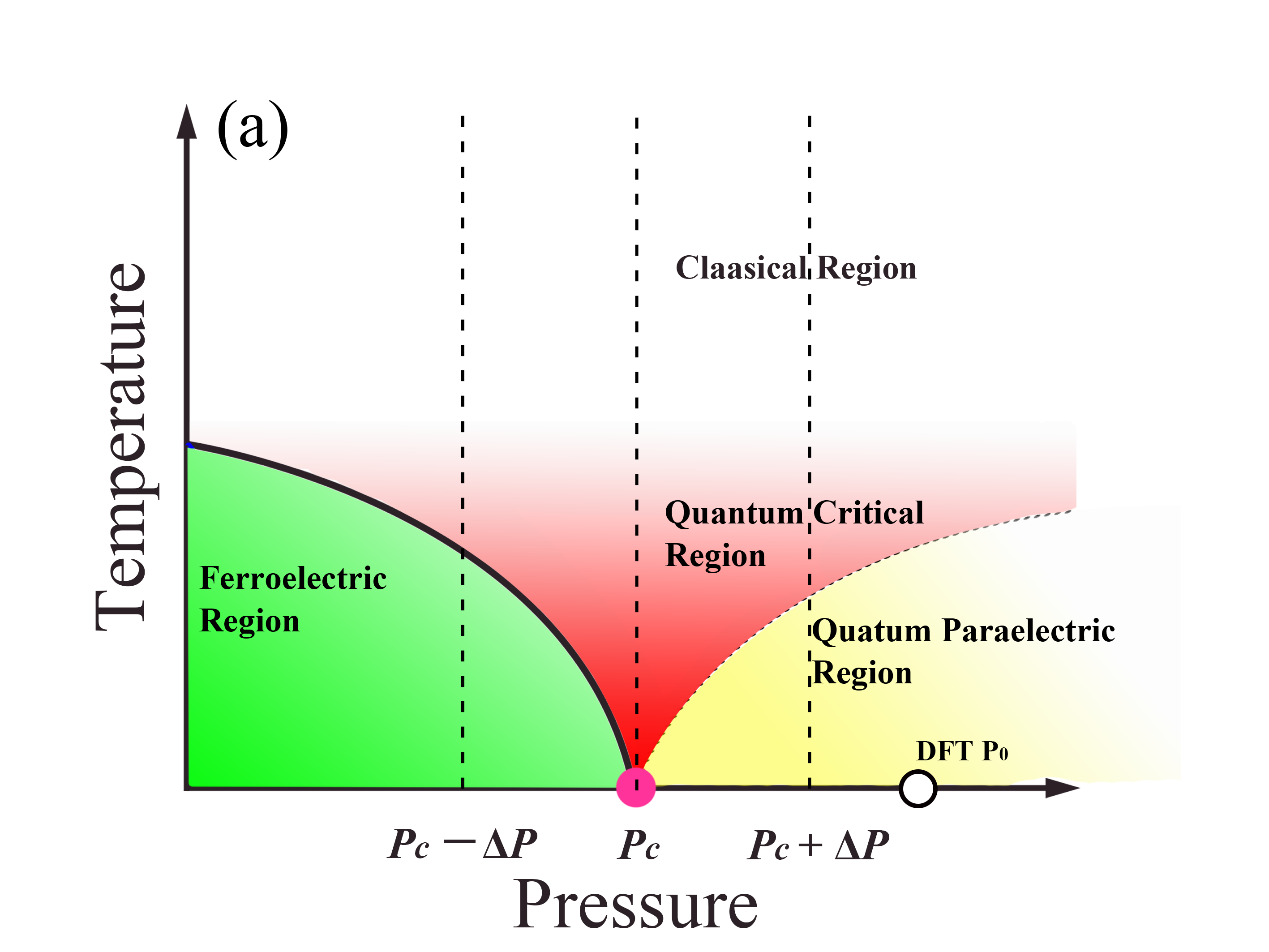}
		\label{PhaseDiagram}
	}
	\quad
	\subfigure{
		\includegraphics[width=\columnwidth]{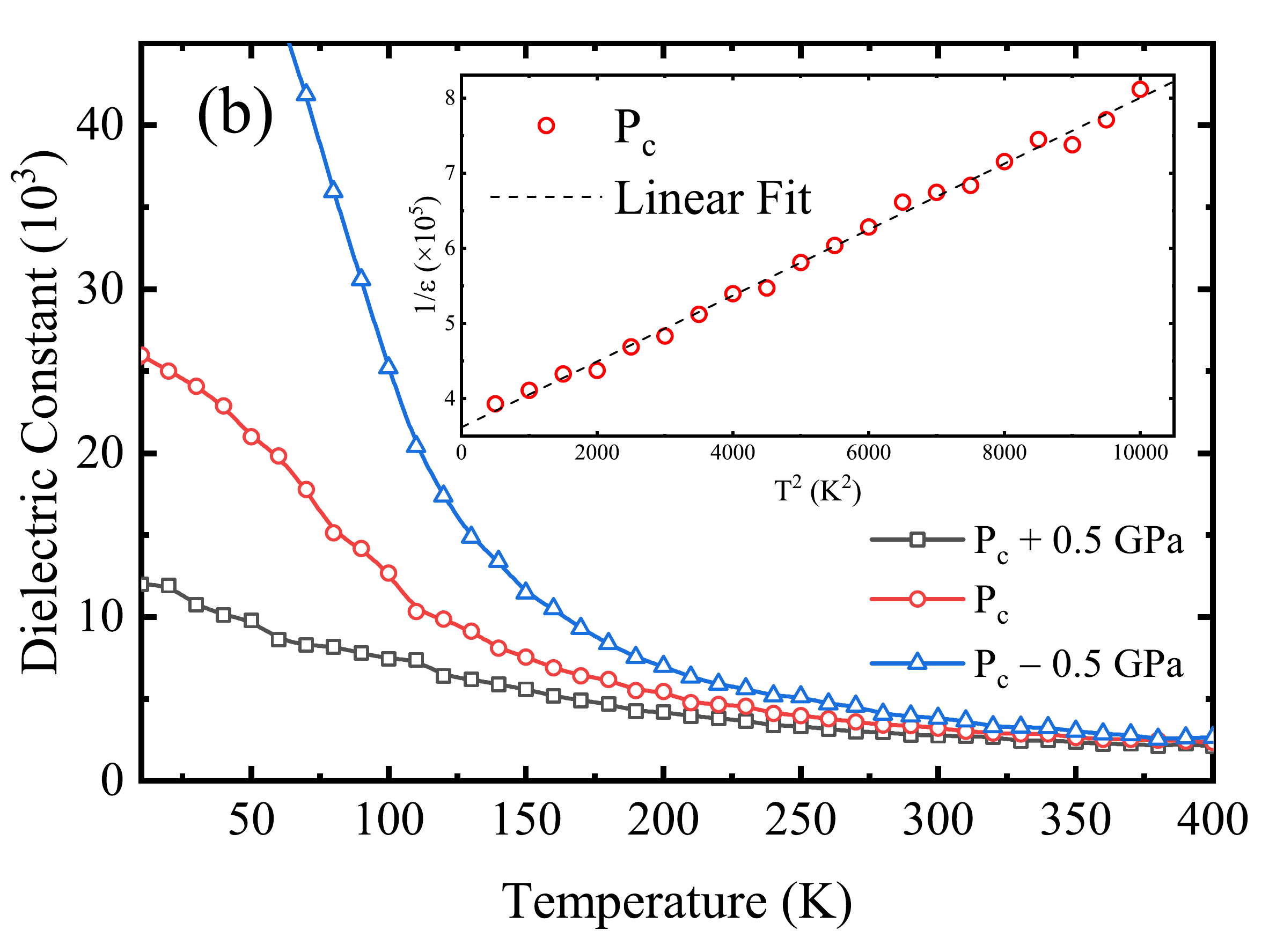}
		\label{QCP}
	}
	\caption{(a) Schematic phase diagram of displacive ferroelectrics based on $\phi^4$ quantum field theory (Ref.~\cite{rowley2014ferroelectric}). (b) DP+QTB simulations calculated dielectric constant near quantum critical point of pressure as a function of temperature.}
\end{figure}

Another property correlated to quantum effects of lattice vibration in SrTiO$_3$ is the quantum behavior of its dielectric constant at low temperature. According to the phase diagram based on $\phi^4$-quantum field model (schematic phase diagram in Figure~\ref{PhaseDiagram}) and experimental result\cite{PhysRevB.19.3593,rowley2014ferroelectric}, SrTiO$_3$ exhibits classical behavior at high temperature, the dielectric constant $\varepsilon(T)$ obey $\frac{1}{T}$ law. While at finite temperature, the dielectric constant $\varepsilon(T)$ exhibit the non-classical $\frac{1}{T^2}$ dependence over a wide range of finite temperature\cite{rowley2014ferroelectric,PhysRevB.19.3593,coak2020quantum,rowley2014ferroelectric}. In order to fairly show the dielectric constant of SrTiO$_3$ calculated by DP+QTB, we first performed the calculation with classical MD as drawn in black in Figure~\ref{ZeroP}. Over the entire temperature range, the $\varepsilon(T)$ calculated by classical MD exhibits the $\frac{1}{T}$ law (inset of Figure~\ref{ZeroP}), which also means that the dielectric constant at zero temperature will diverge.

As for the $\varepsilon(T)$ simulated by DP+QTB (For more accurate results, the number of atoms in the cell of DP+QTB simulations is 135,000. For a detailed discussion, see Supplementary Material S3 and Figure S3\cite{Supplementary}), as drawn in red in Figure~\ref{ZeroP}, it is dramatically different from classical MD. Firstly, it does not diverge near 0K but converges to a constant value. Secondly, its 1/T dependence occurs only above 300K. The possible reason for such huge difference is that the contribution of thermal fluctuations near 0K is almost negligible in classical MD, so the highly collective displacement of atoms brought by the electric field is not suppressed, showing extremely strong electrical polarization. And QTB can successfully introduce quantum fluctuations into MD calculations, this fluctuation will be reflected in the position of the atom, so after time average it can suppress the collective displacement of atomic affected by the electric field. However, the magnitude of dielectric constant calculated by QTB in the Figrue~\ref{ZeroP} is still different from other results\cite{rowley2014ferroelectric,PhysRevB.19.3593,Shin2021,coak2020quantum}, and $T^2$ law is difficult to characterize. As we mentioned, although SrTiO$_3$ is naturally located near the quantum critical point of the displacive ferroelectrics ($P_c$ in Figure~\ref{PhaseDiagram}), due to the accuracy of the DFT exchange correlation function, the lattice constants calculated by DFT is different to experimental results, so DFT simulation result (DFT $P_0$ in Figure~\ref{PhaseDiagram}) is far away from quantum critical point $P_c$. For this problem, we will give a detail discussion later.

\begin{figure}[t]
\centering
\includegraphics[width=\columnwidth]{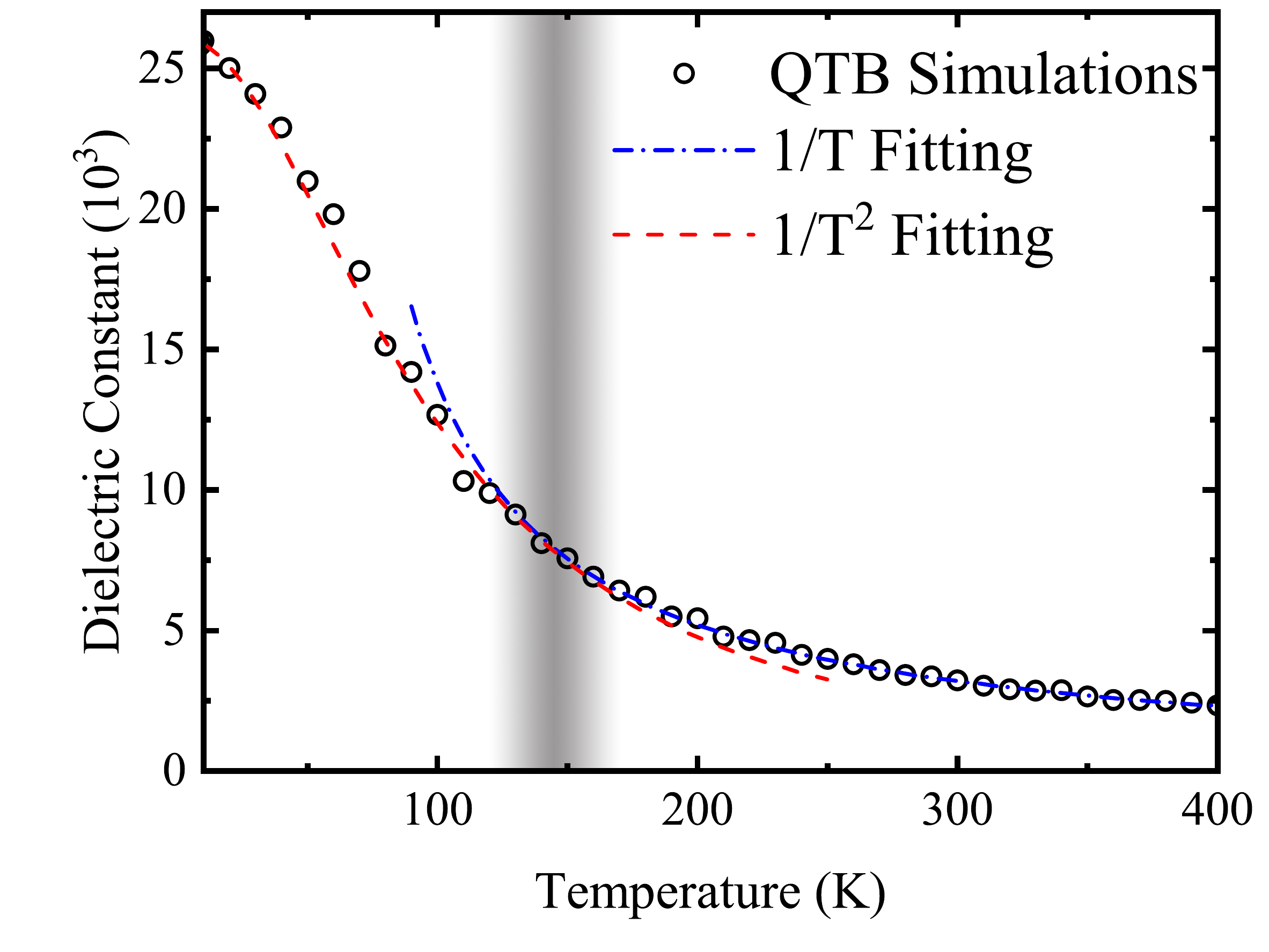}
\caption{DP+QTB simulations calculated dielectric constant at quantum critical point of pressure as a function of temperature.}
\label{QCPfitting}
\end{figure}

So in order to further explore the dielectric constant of SrTiO$_3$ near quantum critical point, we apply different sets of hydrostatic pressure in the DP+QTB simulation to make the pressure close to the quantum critical point ($P_c$ is $\sim$ -4.5GPa), as shown in Figure~\ref{QCP}. According to the phase diagram of SrTiO$_3$ (Figure~\ref{PhaseDiagram})\cite{rowley2014ferroelectric}, the applied hydrostatic pressure can be divided into three categories: $P_c-\Delta P$, $P_c$ and $P_c+\Delta P$. At $P_c$, as the temperature increases from 0 K, the SrTiO$_3$ undergoes the crossover of two phase Regions: Quantum Critical Region-Classical Region. $P_c-\Delta P$ and $P_c+\Delta P$ will undergo Ferroelectric Region and Quatum Paraelectric Region first, respectively.
When a negative hydrostatic pressure is applied relative to the critical pressure $P_c$, as shown in the inset of Figure~\ref{QCP}, $1/\varepsilon(T)$ exhibit the non-classical $\frac{1}{T^2}$ law over a wide range below 100K. This again proves that quantum effect of lattice vibration is not only matter at 0K, but also matters in finite temperature, thus it can not be overlooked in simulations. Then we bring out Figure~\ref{QCPfitting} to inllustrate the phase transition at $P_c$, different types of function fitting are done for different ranges of data. At a wide range of 0 to 150K, $\varepsilon(T)$ exhibits $\frac{1}{T^2}$ law, beyond 150K $\varepsilon(T)$ returns classical behavior and exhibits $\frac{1}{T}$ law. It can be concluded that the quantum fluctuations brought by DP+QTB can effectively decay at high temperature, which indicates that DP+QTB does not bring unphysical results at the classical limit. As for $P_c-\Delta P$, spontaneous ferroelectric polarization occurs at low temperatures, its dielectric constant cannot be represented by the method zone defined in our method. As for $P_c+\Delta P$, a dielectric constant platfrom occurs at low temperatures. For the above two categories, as the temperature increases, the corresponding $\frac{1}{T^2}$ and $\frac{1}{T}$ laws will also appear. At last, we also applied negative pressure to SrTiO$_3$ in classical MD ( see Supplementary Materials S4 and Figure S4 in \cite{Supplementary}), and only $\frac{1}{T}$ law of dielectric constant at all temperatures will occur, which shows that classical MD cannot describe the quantum effect.

\subsection{Discussions}
Now we discuss the influence of anharmonic effect on our conclusion. Our simulation results are carried out by MD simulation at low temperature, and need not involve anharmonic effects. Although the spectral density form of QTB resembles a simple harmonic approximation (eq.\eqref{PSD}), it is actually a general result derived from linear response theory\cite{PhysRev.83.34}. The results of the QTB using the harmonic potential can strictly conform to the quantum theory of the harmonic oscillator, but this does not mean that the application of the anharmonic potential in the QTB is impossible. The results of the combination of anharmonic potential and QTB are also in good agreement with the theoretically and PIMD results\cite{PhysRevLett.107.198902}, suggesting that QTB is capable of handling anharmonic effects, so it is expected to describe the situation of high temperature.

Another question need to be addressed is the discrepancy between DP and experimental results in structure phase transition temperature and DP critical pressure of SrTiO$_3$. First, we have mentioned before it originates from the exchange correlation functionals of DFT. Since the energy difference between the two phases in SrTiO$_3$ is extremely small ($\sim$1 meV), it is extremely difficult for DFT to describe such a subtle energy difference, so there will be discrepancy in the study of problems related to this energy difference. Second, these widely accepted exchange correlation functionals, such as LDA, PBE and PBEsol, they either overestimate or underestimate the experimentally measured lattice constants\cite{Haas2009}. Naively, one would expect the ones that predicts values closest to the experimental results to be the most accurate functional for the studied material. However, such a point of view should be reexamined, as a proper DFT functional intrinsically fails to take into account the zero-point fluctuation, and a ``precise'' functional should result in a systematic underestimation of lattice constants.

\section{conclusions}
We proposed a first principle method DP+QTB which can deal with large scale atomistic systems and quantum effects of lattice vibration. DP+QTB is capable of performing efficient large scale (in both space and time) simulations with DFT accuracy, and thereby solves an outstanding problem in traditional dynamic simulations, which either adopt a classical description that fully ignores quantum effects, or use methods such as path-integral molecular dynamics whose application is limited to small systems. DP+QTB can better predict the phase transition temperature of SrTiO$_3$, and can describe the $\frac{1}{T^2}$ and $\frac{1}{T}$ laws of SrTiO$_3$ dielectric constant in different temperature ranges, where either the quantum or thermal effects dominate. We define $T_{zero}$ to measure the strength of the quantum effect of lattice vibration of a material, and we find that the quantum effect of many systems is far from negligible, which suggest that the corresponding research may expand beyond light elements and extremely low temperatures.

\begin{acknowledgments}
H.-Y. W. thanks Ms. H. L. for help on Figure~\ref{Trans} and Mr. X.-J. Q. for help on Table~\ref{Tzero}. This work was supported by the National Key R$\&$D Program of China (Grants No. 2021YFA0718900), the Key Research Program of Frontier Sciences of CAS (Grant No. ZDBS-LY-SLH008), the
National Nature Science Foundation of China (Grants No. 11974365 and No. 12204496), the K.C. Wong Education Foundation (GJTD-2020-11), and the Science Center of the National Science Foundation of China (52088101).
\end{acknowledgments}

\nocite{*}

\providecommand{\noopsort}[1]{}\providecommand{\singleletter}[1]{#1}%

\end{document}